\documentstyle[12pt]{article}
\def\sp{\phantom{a}}

\makeatletter
\@addtoreset{equation}{section}

\makeatother

\begin{document}
\sloppy
\sloppy
\sloppy

\begin{flushright}{ UT-846 }
\end{flushright}

\vskip 2.5 truecm

\begin{center}
{\Large  11-dimensional curved backgrounds for supermembrane
             in superspace }
\end{center}
\vskip 2 truecm
\centerline{\bf Shibusa Yuuichirou}
\vskip .4 truecm
\centerline {\it Department of Physics,University of Tokyo}
\centerline {\it Hongo 7-3-1,Bunkyo-ku,Tokyo 113-0033,Japan}
\vskip 4 truecm

\begin{abstract}

We compute part of the superfield in terms of the component fields of 
11-dimensional on-shell supergravity by using `Gauge completion' in 
2nd-order formalism. The result is the same as was derived 
recently in 1.5-order formalism by B. de Wit, K. Peeters and J. Plefka. 
We use 2nd-order formalism because in order to hold 
$\kappa $-invariance generally 2nd-order formalism is more hopeful and 
simpler than 1.5-order formalism.
 
\end{abstract}

\newpage

\section{Introduction}
Some years ago, T. Banks, W. Fischler, S. H. Shenker and L. Susskind
(BFSS) proposed
that Matrix theory gives a complete description of light-front 
M-theory~\cite{BFSS}. It had been proposed as a theory of D0-branes by 
E. Witten~\cite{w96}.

In two years following the BFSS conjecture, it has become clear that 
Matrix model encodes a remarkable amount of the structure of M-theory
and 11-dimensional supergravity(for reviews, see~\cite{B98}).
The interaction between gravitons in Matrix theory has been shown to
agree with supergravity to some extent~\cite{oka}.

However, this theory is constructed on flat spacetime, therefore Matrix 
theory on curved backgrounds is required. For single D0-branes, the 
theory on curved backgrounds is expected to be described by 
Born-Infeld action~\cite{lei}. For multi-particle system of D0-branes,
namely Matrix model, the theory on curved backgrounds is as yet unknown.
There are many trials to this problem. For example, starting from flat 
Matrix theory, backgrounds are produced by many D0-branes~\cite{yon}.
The other idea is that it is expected as supermembrane on curved 
backgrounds~\cite{wit98}. In this paper we adopt the later idea.

The theory of supermembrane is described as nonlinear sigma model~\cite{ts87}.
Supermembrane consistently couples to 11-dimensional superspace
backgrounds that satisfy a number of constraints which are equivalent 
to 11-dimensional on-shell supergravity~\cite{kap}. After light cone 
gauge fixing and $\kappa $-symmetry gauge fixing, supermembrane theory 
on flat backgrounds is equivalent to a quantum-mechanical model with 
supersymmetric U(N) gauge symmetry in the large N limit by use of matrix 
regularization~\cite{wit88}. It has a continuous mass spectrum and 
instability~\cite{wit89}, therefore it is expected that supermembrane 
matrix theory describes second quantization of D0-branes~\cite{ts96}.
From the beginning of sigma model, it couples to general backgrounds, 
therefore it is expected that sigma model on curved backgrounds 
is candidate of Matrix theory on curved backgrounds. Actually 
curved backgrounds for supermembrane were investigated~\cite{wit98}.
In this reference, they used `1.5-order formalism'.

In this paper, we use `2nd-order formalism' because in order to hold 
the $\kappa $-invariance generally `2nd-order formalism' is simpler 
and more hopeful than `1.5-order formalism'. We obtain the result 
that is the same as that was drived in this reference~\cite{wit98} up to 
obtained components. And the equations for higher order components 
in `2nd-order
formalism' are different from those in `1.5-order formalism'.    

The paper is organized as follows. In section 2, we review the
supermembrane theory and the condition of $\kappa $-symmetry.
In section 3, we explain our notations of the 11-dimensional 
supergravity and obtain the full algebra of transformations in component 
formalism. In section 4, we explain our notations of the superspace 
geometry and obtain the full algebra of transformations in superspace. In 
section 5, we explain `gauge completion' and compute part of 
the superfields. In section 6, we discuss importance of 2nd-order 
formalism. Other notations and conventions 
used throughout this paper are summarized in Appendix.

\section{Supermembrane theory}
Super membrane theory is described as nonlinear sigma model~\cite{ts87}. It is
written in terms of superspace embedding coordinates 
$Z^M (\xi ) = (X^m (\xi ) , \theta (\xi ))$ , which are functions of the
three world-volume coordinate $\xi ^i (i=0,1,2)$ .

The action is
\begin{eqnarray}
I = \int d^3 \xi (-\frac{1}{2}\sqrt{-g} g^{ij} \Pi _i ^{\sp a}\Pi _j ^{\sp b}
     \eta_{ab} + \frac{1}{2}\sqrt{-g} - \frac{1}{6} \epsilon ^{ijk}
     \Pi _i ^{\sp A} \Pi _j ^{\sp B} \Pi _k ^{\sp C} B_{CBA}), 
\label{mem} 
\end{eqnarray}
where $g_{ij}$ is the metric of the world-volume, $g=det(g_{ij})$ and
$\Pi _i ^{\sp A} \equiv \partial _i Z^M E_M^{\sp A}$.
$E_M^{\sp A}$ is supervielbein, and the 3-form
$B = \frac{1}{6}E^A E^B E^C B_{CBA}$ is potential for the closed 
4-form $H=dB$ .

This action has the following symmetries,

{\it world-volume reparametrization} $\eta ^i(\xi )$
\begin{eqnarray}
  \delta Z^M    &=& \eta ^i \partial _i Z^M , \nonumber \\
  \delta g_{ij} &=& \eta^k \partial _k g_{ij} + 2\partial _{(i}\eta ^k 
                  g_{j)k},
\end{eqnarray}
{\it \sp \sp \sp $\kappa $-symmetry $\kappa ^{\alpha }(\xi )$}
\begin{eqnarray}
  \delta Z^M E_M ^{\sp a}       &=& 0,  \nonumber \\
  \delta Z^M E_M ^{\sp \alpha}  &=& (1+\Gamma ^{\alpha }_{\sp \beta })
                                    \kappa ^{\beta }, \nonumber \\
  \delta (\sqrt{-g}g^{ij})      &=& -2(1+\Gamma ^{\alpha }_{\sp \beta})
                                    \kappa ^{\beta }\Gamma_{ab \sp \alpha
                                      \gamma }\Pi _n^{\sp \gamma }g^{n(i}
                                      \epsilon ^{j)kl}\Pi_k^{\sp a}
                                      \Pi _l^{\sp b} \nonumber \\
                                & & +\frac{-2}{3\sqrt{-g}}
                                      \kappa ^{\alpha }\Gamma _{c \sp 
                                      \alpha \beta }\Pi ^{k \beta }\Pi _k
                                      ^c \epsilon ^{mn(i} \epsilon ^{j)
                                      pq} \nonumber \\
                                & & (\Pi _m^{\sp a}\Pi _{pa}\Pi _n^{
                                      \sp b}\Pi _{qb} +\Pi_m^{\sp a}\Pi _{pa}
                                      g_{nq}+g_{mp}g_{nq}),
\end{eqnarray}
where $\kappa ^{\alpha }(\xi )$ is anticommuting space time spinor and 
the matrix 
$\Gamma $ is defined by
\begin{eqnarray}
 \Gamma = \frac{1}{6 \sqrt{-g}}\epsilon ^{ijk}\Pi _i^{\sp a}
          \Pi _j^{\sp b} \Pi _k^{\sp c}\Gamma _{abc}.  
\end{eqnarray}
Up to surface terms the $\kappa $-invariance of this action imposes the 
following constraints on the 11-dimensional superspace 
geometry~\cite{kap}.
\begin{eqnarray}
 T^a_{\sp \alpha \beta}            &=& -2 \Gamma ^a _{\sp \sp 
                                       \alpha \beta}, \nonumber \\
 H_{\alpha \beta\ ab}              &=& -2 \Gamma _{ab \sp 
                                       \alpha \beta}, \nonumber \\
 H_{\alpha \beta \gamma \delta }=
 H_{\alpha \beta \gamma d}=
 H_{\alpha bcd}                    &=& 0, \nonumber \\
 T^{\alpha }_{\sp \beta \gamma}=
 T^a_{\sp bc}=T^a_{\sp b \gamma }  &=& 0.
\label{kapcon}
\end{eqnarray}
The equations of motion which follow from this action are 
\begin{eqnarray}
\label{eom}
  0 &=& g _{ij} - \Pi _i^{\sp a}\Pi _j^{\sp b} \eta _{ab}, \\
  0 &=& \partial _i (\sqrt{-g} g^{ij}\Pi _j^{\sp a})+\sqrt{-g} \Pi _i^{
        \sp b}\Pi ^{iC} \Omega _{Cb}^{\sp \sp a} \nonumber \\
    & &  +\epsilon^{ijk}\Pi _{ib}(\Pi _j^{\sp \alpha}\Gamma ^{ab}_{\sp \sp 
        \alpha \beta}\Pi _k ^{\sp \beta}) \nonumber \\
    & & + \frac{1}{6}\epsilon^{ijk}\Pi _i^{\sp b}\Pi _j^{\sp c}
        \Pi _k^{\sp d}H^a_{\sp bcd},  \\
  0 &=& ((1- \Gamma)\Pi ^{ia}\Gamma _a)^{\alpha}_{\sp \beta}\Pi _i^{\sp \beta}.
\end{eqnarray}
Using (\ref{eom}), the action (\ref{mem}) can be rewritten as
\begin{eqnarray}
 I = \int d^3 \xi (-\sqrt{-g}-\frac{1}{6}\epsilon ^{ijk}\Pi _i^{\sp A}
      \Pi _j^{\sp B}\Pi _k^{\sp C}B_{CBA}).
\label{mem2}
\end{eqnarray}
Flat superspace is characterized by
\begin{eqnarray}
  E_m^{\sp a}             &=& \delta _m ^{\sp a}, \nonumber \\
  E_{\mu }^{\sp a}        &=& -(\Gamma ^a \theta )_{\mu }, \nonumber \\
  E_m^{\sp \alpha }       &=& 0,\nonumber \\
  E_{\mu }^{\sp \alpha }  &=& \delta _{\mu }^{\sp \alpha }, \nonumber \\
 B_{\gamma \beta \alpha } &=& \bar{\theta }\Gamma _{
                              mn(\gamma }\bar{\theta }
                              \Gamma ^m_{\sp \beta }
                              \bar{\theta }\Gamma ^n
                              _{\sp \alpha )}, \nonumber \\
 B_{c \beta \alpha }      &=& -\bar{\theta }\Gamma _{
                              cd(\beta }\bar{\theta }
                              \Gamma ^d_{\sp \alpha ) }, 
                              \nonumber \\
 B_{cb\alpha }            &=& (\bar{\theta }\Gamma _{cb}
                              )_{\alpha },
                              \nonumber \\
 B_{cba}                  &=& 0.
\label{flat}
\end{eqnarray}
In flat superspace, the action (\ref{mem2}) becomes
\begin{eqnarray}
 I=\int d^3\xi \{ -\sqrt{-g}-\epsilon ^{ijk}\bar{\theta }\Gamma _{mn}
               \partial _k \theta(\frac{1}{2}\partial _i X^m(\partial _j X^n +
               \bar{\theta }\Gamma ^n \partial _j \theta ) 
               + \frac{1}{6}\bar{\theta }\Gamma ^m \partial _i \theta 
               \bar{\theta }\Gamma ^n \partial _j\theta )\} .
\label{mem3}
\end{eqnarray}
From this action, after gauge fixing of $\kappa $-symmetry and
reparametrization invariance, using matrix regularization we obtain Matrix 
model~\cite{wit88}. In this paper we don't struggle in more detail.

\newpage
\section{11-dimensional supergravity}
Supergravity in 11-dimensional spacetime is based on 
`elfbein' field $e_m^{\sp a}$, a Majorana gravitino field 
$\psi _m^{\sp \alpha }$ and third rank antisymmetric gauge field $C_{klm}$.
Its Lagrangian can be written as follows~\cite{CJ78}\cite{wit98}.
\begin{eqnarray}
 L &=& -\frac{1}{2}eR -2e\bar{\psi }_m \Gamma ^{mnl}D_n(\frac{1}{2}(\omega + 
       \hat{\omega }))\psi _l - \frac{1}{96}eF^2 \nonumber \\
   & & -\frac{1}{41472}\epsilon ^{m_1 ... m_{11}}F_{m_1 ... m_4}
       F_{m_5 ... m_8}C_{m_9 ... m_{11}} \nonumber \\
   & & -\frac{1}{96}e(\bar{\psi }_n \Gamma ^{m_1 ... m_4 nl}\psi _l +
        12\bar{\psi }^{m_1}\Gamma ^{m_2 m_3}\psi ^{m_4})(F+\hat{F})_
        {m_1 ...m_4}.
\end{eqnarray}
where $e = det e_m^{\sp a}$, and $\omega _{m\sp b}^{\sp a}$denotes the spin 
connection
\begin{eqnarray}
 \omega _{m\sp b}^{\sp a} &=& -e^{na}\partial _{[m}e_{n]b}+e^{la}e^n_{\sp b}
                              e_m^{\sp c}\partial _{[l}e_{n]c}+e^n_{\sp b}
                              \partial _{[m}e_{n]}^{\sp a} \nonumber \\
                          & & +2(\bar{\psi _m}\Gamma _b \psi ^a
                                 +\bar{\psi _b}\Gamma _m \psi ^a
                                 -\bar{\psi _m}\Gamma ^a \psi _b)-\frac{1}{2}
                                 \bar{\psi _n}\Gamma ^{\sp a \sp np}_
                                 {m \sp b}\psi _p ,
\label{conn}
\end{eqnarray}
and $F_{klmn}$ denotes the field strength of the antisymmetric tensor
\begin{eqnarray}
  F_{klmn}=4\partial _{[k}C_{lmn]}.
\end{eqnarray}
The derivative $D_m $ is covariant with respect to local Lorentz 
transformations,
\begin{eqnarray}
  D_m(\omega )\epsilon \equiv (\partial _m - \frac{1}{4}\omega _{mab}
                               \Gamma ^{ab})\epsilon.
\end{eqnarray}
Supersymmetry transformations are equal to
\begin{eqnarray}
\label{st}
  \delta _s e_m ^{\sp a} &=      & 2 \bar{\epsilon }\Gamma ^a \psi _m, 
                                  \nonumber \\
  \delta _s \psi _m      &=      & D_m(\hat{\omega })\epsilon + 
                                  T_m ^{\sp rstu}\epsilon \hat{F}_{rstu} 
                                  \equiv \hat{D}\epsilon, \nonumber \\
  \delta _s C_{klm}      &=      & -6 \bar{\epsilon }\Gamma _{[kl}\psi _{m]},\\
  \mbox{with} ,  
  T_m ^{\sp rstu}        &\equiv & \frac{1}{288}(\Gamma _m^{\sp rstu}-8
                                     \delta _m^{[r}\Gamma ^{stu]}),
\end{eqnarray}
where $\hat{F}$ is the supercovariant field strength,
\begin{eqnarray}
 \hat{F}_{klmn} = F_{klmn}+12 \bar{\psi}_{[k}\Gamma _{lm}\psi _{n]},
\end{eqnarray}
and $\hat{\omega }$ is the supercovariant spin connection
\begin{eqnarray}
 \hat{\omega }_{m \sp b}^{\sp a} = \omega _{m \sp b}^{\sp a}+ 
                                   \frac{1}{2}\bar{\psi _n}\Gamma ^
                                   {\sp a \sp np}_{m \sp b}\psi _p.  
\end{eqnarray}
Note that the spin connection $\omega $ has supersymmetry variation
according to elfbein and gravitino's variation in 2nd-order 
formalism~\cite{des76}. While in 1.5-order formalism, it is defined
as a dependent field determined by its equation of motion, whereas its 
supersymmetry variation is treated as if it were an independent 
field~\cite{wes77}. In this paper we use 2nd-order formalism.

The gauge transformations are equal to 
\begin{eqnarray}
\label{ct}
  \delta _c C_{mnl} = 3 \partial _{[m} \xi_{nl]}.
\end{eqnarray}
The local Lorentz transformations are equal to
\begin{eqnarray}
\label{lt}
  \delta _l e_m^{\sp a}  &=& \lambda ^a_{\sp b}e_m ^{\sp b}, \nonumber \\
  \delta _l \psi 
   _m^{\sp \alpha}       &=& \frac{1}{4}\lambda_{ab}\Gamma^{ab\alpha }_{
                              \sp \sp \sp \beta}\psi _m ^{\sp \beta },
                              \nonumber \\
  \delta_l \omega 
   ^{\sp a}_{m\sp b}     &=& \partial _m \lambda^a_{\sp b}+\lambda^a_
                              {\sp c}\omega _{m\sp b}^{\sp c}- \omega 
                             _{m\sp c}^{\sp a}\lambda ^c_{\sp b}.
\end{eqnarray}
The general coordinate transformation are equal to 
\begin{eqnarray}
\label{gt}
 \delta _g e_m^{\sp a}    &=& \xi^n\partial _n e_m^{\sp a}+\partial _m
                              \xi ^n e_n ^{\sp a}, \nonumber \\
 \delta _g\omega _{m
   \sp b}^{\sp a}         &=& \xi^n\partial _n \omega _{m\sp b}^{\sp a}
                              +\partial _m\xi ^n \omega _{n\sp b}^{
                              \sp a}, \nonumber \\
 \delta _g \psi _m^
  {\sp a}                 &=& \xi^n\partial _n \psi _m^{\sp a}+\partial _m
                              \xi ^n \psi _n ^{\sp a}, \nonumber \\
 \delta _g C_{mnl}        &=& \xi^n\partial _n C_{mnl}+3\partial _{[m}
                              \xi ^k C_{|k|nl]}.
\end{eqnarray}

We obtain the full algebra of these transformations as follows
\begin{eqnarray}
 [\delta _g(\xi _1)+\delta _s(\epsilon _1)+\delta _l(\lambda _1)+\delta _c
    (\xi_{1mn}),\delta _g(\xi _2)+\delta _s(\epsilon _2)+\delta _l(\lambda 
   _2)+\delta _c(\xi_{2mn})]  \nonumber \\
 = \delta _g(\xi _3)+\delta _s(\epsilon _3)+
    \delta _l(\lambda _3)+\delta _c(\xi_{3mn}),
\end{eqnarray}
where 
\begin{eqnarray}
\label{alg}
 \xi _3^m     &=& \xi _2^n \partial _n \xi _1^m + \bar{\epsilon }_2\Gamma ^m
                  \epsilon _1 -(1 \leftrightarrow 2), \nonumber \\
 \epsilon _3  &=& -\bar{\epsilon }_2\Gamma ^n\epsilon _1 \psi _n - \xi _1^n
                  \partial _n \epsilon_2 + \frac{1}{4}\lambda_{2cd}\Gamma^{
                   cd}\epsilon _1 -(1 \leftrightarrow 2), \nonumber \\
 \lambda _
 {3\sp b}^{
  \sp a}     &=& -\bar{\epsilon }_2\Gamma ^n\epsilon _1 \hat{\omega }_{n\sp 
                 b}^{\sp a}-\xi _1^n \partial _n\lambda _{2\sp b}^{\sp a} +
                 \lambda _{2\sp c}^{\sp a}\lambda _{1\sp b}^{\sp c} \nonumber 
                 \\
             & & +\frac{1}{144}\bar{\epsilon }_2(\Gamma _{\sp b}^{a\sp rstu}
                 \hat{F}_{rstu}+24\Gamma _{rs}\hat{F}_{\sp b}^{a \sp rs})
                 \epsilon_1 -(1 \leftrightarrow 2), \nonumber \\
 \xi _{3mn}  &=& -\bar{\epsilon }_2\Gamma ^k\epsilon _1 C_{kmn}
                 -\bar{\epsilon }_2\Gamma _{mn}\epsilon _1-\xi _1^k \partial 
                 _k \xi _{2mn}-2\xi _1^k\partial_{[m}\xi _{2n]k} \nonumber \\
             & & -(1 \leftrightarrow 2). 
\end{eqnarray}

\section{Superspace representation}
In this section, we explain notations of the superspace geometry and 
obtain the full algebra of transformations in 11-dimensional superspace. 
As usual, we suppose
that the superspace has Lorentzian tangent space structure and 
the vielbein $E_M^{\sp A}$ and the connection
 $\Omega _{MA}^{\sp \sp \sp B}$ and the corresponding 1-forms
\begin{eqnarray}
 E^A               &=& dz^M E_M^{\sp A}, \nonumber \\
 \Omega _A^{\sp B} &=& dz^M \Omega _{MA}^{\sp \sp \sp B}.
\end{eqnarray}
The Lorentzian assumption implies 
\begin{eqnarray}
 \Omega _{ab}            &=& -\Omega _{ba},  \nonumber \\
 \Omega _{\alpha b}      &=& 0,              \nonumber \\
 \Omega _{\alpha \beta } &=& \frac{1}{4}\Omega _{ab}\Gamma ^{ab}_{\sp \sp \sp
                              \alpha \beta }.
\end{eqnarray}
There are also 3-form potential 
\begin{eqnarray}
 B    =  \frac{1}{3!}dz^L dz^M dz^N  B_{NML}, 
\end{eqnarray}
and field strength 4-form
\begin{eqnarray}
 H =  dB  =  \frac{1}{4!}dz^K dz^L dz^M dz^N  H_{NMLK}. 
\end{eqnarray}
From these basic fields we can define the torsion and curvature as
follows
\begin{eqnarray}
\label{tor}
  T^A          &\equiv & DE^A, \nonumber \\
  R_A^{\sp B}  &\equiv & d\Omega _A ^{\sp B}+ \Omega _A^{\sp C}
                         \Omega _C^{\sp B}, 
\end{eqnarray}
where covariant derivative $D$ is defined as follows
\begin{eqnarray}
 DE^A \equiv dE^A + E^B \Omega_B^{\sp A}.
\end{eqnarray}
One then has the Bianchi identities
\begin{eqnarray}
 DT^A         &=& E^B R_B^{\sp A}, \nonumber \\
 DR_A^{\sp B} &=& 0, \nonumber \\
 DH           &=& 0 .
\end{eqnarray}
The supertransformation is equal to 
\begin{eqnarray}
\label{TT}
 \delta _T X_{M_p ... M_1} = \Xi ^K \partial _K X_{M_p ... M_1}+p\partial 
                             _{[M_p} \Xi ^K X_{|K|M_{p-1}...M_1]}
\end{eqnarray}
for p-form's components.
The local Lorentz transformations are equal to 
\begin{eqnarray}
\label{LT}
 \delta _L E^A              &=& E^B \Lambda _B^{\sp A},  \nonumber \\
 \delta _L \Omega_B^{\sp A} &=& -\Lambda_B^{\sp C}\Omega _C^{\sp A} +
                                 \Omega _B^{\sp C}\Lambda _C^{\sp A}-
                                d\Lambda _B ^{\sp A} .
\end{eqnarray}
The supergauge transformations are equal to 
\begin{eqnarray}
\label{GT}
  \delta _G B_{LMN} = 3\partial _{[L}\Xi _{MN]} .
\end{eqnarray}
We obtain the full algebra of these transformations as follows
\begin{eqnarray}
 [\delta _T(\Xi _1)+\delta _L(\Lambda _1)+\delta _G
    (\Xi_{1MN}),\delta _T(\Xi _2)+\delta _L(\Lambda 
   _2)+\delta _G(\Xi_{2MN})]  \nonumber \\
    = \delta _T(\Xi _3)+
    \delta _L(\Lambda _3)+\delta _G(\Xi_{3MN}),
\end{eqnarray}
where,
\begin{eqnarray}
\label{ALG}
 \Xi _3^K                   &=& \Xi _2^L\partial _L \Xi _1^K +\delta _1 
                                \Xi _2^K-(1 \leftrightarrow 2), \nonumber \\
 \Lambda _{3A}^{\sp \sp B}  &=& -\Xi _1^K \partial _k\Lambda _{2A}^
                                {\sp \sp B} +\delta _1\Lambda _{2A}^
                                {\sp \sp B} +\Lambda_{1A}^{\sp \sp C}
                                \Lambda _{2C}^{\sp \sp B}
                                 -(1 \leftrightarrow 2), \nonumber \\
 \Xi _{3MN}                 &=& \delta_1\Lambda _{2MN}-\Xi _1^K\partial _K
                                \Xi _{2MN} -2\partial _{[M}\Xi _{2N]K}
                                \Xi _1^K -(1 \leftrightarrow 2).          
\end{eqnarray}

There are a great number of component fields in superspace. Thus if we try
to identify superspace representation as ordinary supergravity,
there are a great number of unknown degrees of freedom. The method of this
identification is known as `gauge completion'~\cite{gc}. We shall  
explain it in the next section.

\section{Gauge completion}
`Gauge completion' was introduced in order to identify superspace 
representation as on-shell supergravity~\cite{gc}. In this section we
review this method and compute part of the superfield in terms of 
the on-shell supergravity fields.

Using this method in 2nd-order formalism, up to first order in 
anticommuting coordinates, the superfield component was investigated 
by E. Cremmer and S. Ferrara~\cite{kap}. And in 1.5-order formalism,  
part of component at second order in anticommuting coordinates was 
investigated by B. de Wit, K. Peeters and J. Plefka~\cite{wit98}.

`Gauge completion' is finding the superfields and superparameters which 
are compatible with ordinary supergravity. That is to say,
supertransformations (\ref{TT}) - (\ref{GT}) are identified as 
transformations in 11-dimensional spacetime (\ref{st}) - (\ref{gt}) and 
the $\theta =0$ components of superfields and super parameters are 
identified as the fields and parameters of ordinary supergravity.

Firstly, a gauge is chosen as follows
\begin{eqnarray}
 E_m^{\sp a(0)}         &=& e_m^{\sp a}, \nonumber \\
 E_m^{\sp \alpha (0)}   &=& \psi _m^{\sp \alpha }, \nonumber \\
 \Omega _{mb}^
 {\sp \sp \sp a(0)}       &=& -\hat{\omega }_{m \sp b}^{\sp a}, \nonumber \\
 \Xi ^{m(0)}            &=& \xi ^m, \nonumber \\
 \Xi ^{\alpha (0)}      &=& \epsilon ^{\alpha }, \nonumber \\
 \Xi ^{(0)}_{mn}        &=& \xi _{mn}, \nonumber \\
 B_{mnl}^{(0)}          &=& C_{mnl}.  
\end{eqnarray}
From (\ref{LT}) and (\ref{lt}), we obtain 
\begin{eqnarray}
 \Lambda _b ^{\sp a \sp (0)} = \lambda _{\sp b}^a .
\end{eqnarray}
And we introduce some assumptions as follows
\begin{eqnarray}
 \Xi _{\mu N}^{(0)}    &=& 0, \nonumber \\
 \Xi _{\mu N}^{(1)}    &=& 0 . 
\end{eqnarray}
Then, the higher order component in anticommuting coordinates can be
obtained  by requiring consistency between the algebra of 
superspace supergravity and that of ordinary supergravity.

To make this procedure clear, we write one simple example explicitly.
We take $\theta =0 $ component of vielbein.

According to superspace algebra,
\begin{eqnarray}
 \delta E_m^{\sp a}
 |_{\theta =0}         &=& (\Xi ^K\partial _KE_m^{\sp a}
                           +\partial _m \Xi ^K E_K^{\sp a}+ E_m^{\sp b}\Lambda 
                           _b^{\sp a})|_{\theta =0} \nonumber \\
                       &=& \xi ^k\partial _k e_m^{\sp a} + \partial _m
                           \xi ^ke_k^{\sp a}+\epsilon ^{\nu }\partial _{\nu }
                           (E_m^{\sp a (1)})\nonumber \\
                       & & +\partial_m\epsilon ^{\nu }
                           E_{\nu }^{\sp a(1)}+e_m^{\sp b}\lambda ^a_{\sp b},  
\end{eqnarray}
while in ordinary supergravity
\begin{eqnarray}
 \delta e_m^{\sp a} = \xi ^k\partial _k e_m^{\sp a}+\partial _m \xi ^k 
                      e_k^{\sp a}+2\bar{\epsilon }\Gamma ^a \psi _m + 
                      \lambda ^a_{\sp b}e_m^{\sp b}.
\end{eqnarray}
Thus, we obtain 
\begin{eqnarray}
 E_{\nu }^{\sp a \sp (0)}  &=& 0, \nonumber \\
 E_m^{\sp a (1)}           &=& 2\bar{\theta }\Gamma ^a\psi _m .
\end{eqnarray}
By this procedure, we obtain the following results.
\begin{eqnarray}
 E_m^{\sp a}   &=& e_m^{\sp a}+2\bar{\theta }\Gamma ^a\psi _m -\frac{1}{4}
                   \bar{\theta }\Gamma^{acd}\theta \hat{\omega }_{mcd}+
                   \frac{1}{72}\bar{\theta }\Gamma _m^{\sp rst}\theta \hat{F}
                   _{rst}^{\sp \sp \sp a} \nonumber \\
               & & +\frac{1}{288}\bar{\theta }\Gamma ^{rstu}\theta \hat{F}
                   _{rstu}e_m^{\sp a}-\frac{1}{36}\bar{\theta }\Gamma 
                   ^{astu}\theta \hat{F}_{mstu}+{\cal O}(\theta ^3), \\
 E_m^
 {\sp \alpha}  &=& \psi _m^{\sp \alpha}-\frac{1}{4}\hat{\omega }_{mab}(\Gamma 
                   ^{ab}\theta )^{\alpha }+(T_m^{\sp rstu} \theta )^{\alpha}
                   \hat{F}_{rstu}+{\cal O}(\theta ^2), \\
 E_{\mu }^{
 \sp a}        &=& -(\Gamma ^a \theta)_{\mu }+{\cal O}(\theta ^2),\\
 E_{\mu }^{
 \sp \alpha }  &=& \delta _{\mu }^{\sp \alpha }+{\cal O}(\theta ^2), \\
 \Omega _
 {\mu b}^{
 \sp \sp a}    &=& \frac{1}{144}\{(\Gamma ^{a\sp rstu}_{\sp b}\theta )_{\mu }
                   \hat{F}_{rstu}+24(\Gamma _{rs}\theta )_{\mu }\hat{F}_
                   {\sp b}^{a \sp rs} \} +{\cal O}(\theta ^2),  \\
 \Omega _{mab} &=& \hat{\omega }_{mab}+2\bar{\theta }\{ e^n_{\sp a}e^k_
                   {\sp b}(-\Gamma _kD_{[m}\psi _{n]}+\Gamma _nD_{[m}
                   \psi _{k]}+\Gamma _mD_{[n}\psi _{k]})\} \nonumber \\
               & & -\bar{\psi }_a\Gamma _bT_m^{\sp rstu}\theta \hat{F}_{rstu}
                   +\bar{\psi }_m\Gamma _bT_a^{\sp rstu}\theta \hat{F}_{rstu}
                   +\bar{\psi }_b\Gamma _aT_m^{\sp rstu}\theta \hat{F}_{rstu}
                   \nonumber \\
               & & -\bar{\psi }_m\Gamma _aT_b^{\sp rstu}\theta \hat{F}_{rstu}
                   -\bar{\psi }_m\Gamma _aT_b^{\sp rstu}\theta \hat{F}_{rstu}
                   \nonumber \\
               & & +\bar{\psi }_b\Gamma _mT_a^{\sp rstu}\theta \hat{F}_{rstu}
                   +{\cal O}(\theta ^2),  \\
 B_{mnl}       &=& C_{mnl}-6\bar{\theta }\Gamma _{[mn}\psi _{l]}+\frac{3}{4}
                   \hat{\omega }_{[l}^{\sp \sp cd}\bar{\theta }\Gamma _{
                   mn]cd}\theta -\frac{3}{2}\hat{\omega }_{[lmn]}\theta ^2
                   \nonumber \\
               & & -\frac{1}{96}\bar{\theta }\Gamma _{mnl}^{\sp \sp \sp rstu}
                   \theta \hat{F}_{rstu}-\frac{3}{8}\bar{\theta }\Gamma _{
                   [l}^{\sp \sp rs}\theta \hat{F}_{|rs|mn]}-12\bar{\theta }
                   \Gamma _a\psi _{[m}\bar{\theta }\Gamma ^a_{\sp n}\psi _{l]}
                   \nonumber \\
               & & +{\cal O}(\theta ^3), \\
 B_{lm\mu }    &=& (\bar{\theta }\Gamma _{lm})_{\mu }+{\cal O}(\theta ^2), \\
 B_{m\mu \nu } &=& (\bar{\theta }\Gamma _{mn})_{(\mu }(\bar{\theta }\Gamma ^n
                   )_{\nu )}+{\cal O}(\theta ^2), \\
 B_{\mu \nu 
 \rho }        &=& (\bar{\theta }\Gamma _{mn})_{(\mu }(\bar{\theta }\Gamma ^m
                   )_{\nu }(\bar{\theta }\Gamma ^n)_{\rho )}
                   +{\cal O}(\theta ^2), \\
 \Xi ^m        &=& \xi ^m +\bar{\theta }\Gamma ^m\epsilon -\bar{\theta }
                   \Gamma ^n\epsilon \bar{\theta }\Gamma ^m\psi _n
                   +{\cal O}(\theta ^2), \\
 \Xi ^{\mu}    &=& \epsilon ^{\mu }-\frac{1}{4}\lambda _{cd}(\Gamma ^{cd}
                   \theta )^{\mu }-\bar{\theta }\Gamma ^n\epsilon 
                   \psi _n ^{\sp \mu }+{\cal O}(\theta ^2), \\
 \Lambda _b^{
  \sp a}       &=& \lambda ^a_{\sp b}-\bar{\theta }\Gamma ^n\epsilon 
                   \hat{\omega }_{n \sp b}^{\sp a}+\frac{1}{144}\bar{\theta }
                   (\Gamma ^{a\sp rstu}_{\sp b}\hat{F}_{rstu}+24\Gamma _{rs}
                   \hat{F}_{\sp b}^{a \sp rs})\epsilon
                   +{\cal O}(\theta ^2), \\
 \Xi _{mn}     &=& \xi _{mn}-(\bar{\theta }\Gamma ^p\epsilon C_{pmn}+
                   \bar{\theta }\Gamma _{mn}\epsilon ) 
                   +{\cal O}(\theta ^2), \\
 \Xi _{m\mu }  &=& {\cal O}(\theta ^2), \\
 \Xi _{\mu 
        \nu }  &=& {\cal O}(\theta ^2).
\end{eqnarray}
3-form fields are obtained up to first order in anticommuting coordinates. In
order to compute this at second order in anticommuting coordinates, the 
superparameter $\Xi _{MN}$ at second order is needed. And in
order to compute $\Xi _{MN}$  at second order in anticommuting
coordinates, 
the superparameter $\Xi ^M$ at second order is needed. Thus we can't compute 
3-form fields at second order. However, because flat geometry is known 
(\ref{flat}), we include the $\theta ^3$ term in $B_{\mu \nu \rho }$ and 
the $\theta ^2$ term in $B_{m\mu \nu }$ for completeness.

We obtained all components which was required in order to write the action 
(\ref{mem}) up to $\theta ^2$ term. 

The components of vielbein and 3-form field obtained above is the same as was 
constructed before~\cite{wit98}. Thus this result holds the invariance 
of $\kappa $-symmetry. However the other components and the equations 
which components of vielbein and 3-form and superparameter at second order 
in anticommuting coordinates must obey are different from those in 
reference~\cite{wit98}.
These can be written explicitly as follows, 
\begin{eqnarray}
0 &=& X^k \partial _ke_m^{\sp a}+\partial _mX^k e_k^{\sp a}
      -\bar{\epsilon _2 }\Gamma ^k\epsilon _1(\psi _k^{\sp \mu }
      \partial _{\mu }E_m^{\sp a(2)}+\hat{\omega }_{[m|cd|}\bar{\theta }
      \Gamma ^a\Gamma ^{cd}\psi _{k]}) \nonumber \\
  & & +2\bar{\theta }\Gamma ^b \psi _m \{ -\bar{\epsilon }_2\Gamma ^n
      \epsilon _1 \hat{\omega }_{n\sp b}^{\sp a}+\frac{1}{144}
      \bar{\epsilon }_2 (\Gamma ^{a\sp rstu}_{\sp b}\hat{F}_{rstu}+
      24\Gamma _{rs}\hat{F}_{\sp b}^{a\sp rs})\epsilon _1 \}
      \nonumber \\
  & & -\frac{1}{288}\bar{\epsilon }_2 (\Gamma ^{\sp \sp rstu}_{cd}\hat{F}_
      {rstu}+24\Gamma _{rs}\hat{F}_{cd}^{\sp \sp rs})\epsilon _1
      \bar{\theta }\Gamma ^a\Gamma ^{cd}\psi _m 
      + e_m^{\sp b}Y_b^{\sp a}  \nonumber \\
  & & +2\{ \bar{\epsilon }_2^{\sp \nu }\partial _{\nu }\Xi _1^{
       \sp \mu (2)}-\bar{\theta }\Gamma ^k \epsilon _2\bar{\psi }_k
       \Gamma ^n\epsilon _1\psi _n^{\sp \mu}+\frac{1}{4}
       \bar{\theta }\Gamma ^k \epsilon _2\hat{\omega }_{kcd}(\Gamma ^{cd}
       \epsilon _1)^{\mu} \nonumber \\
  & & -\bar{\theta }\Gamma ^k \epsilon _2 
       \hat{F}_{rstu}(T_k^{\sp rstu}\epsilon _1)^{\mu } \} 
       (\Gamma ^a\psi_m)_{\mu }  \nonumber \\ 
  & &  - (1 \leftrightarrow 2),
\end{eqnarray}
where 
\begin{eqnarray}
 X^k         &=& \epsilon _2^{\nu }\partial _{\nu }\Xi _1^{k(2)}+\bar{\theta }
                 \Gamma ^n\epsilon _1 \bar{\epsilon }_2\Gamma ^k\psi _n +
                 \bar{\epsilon }_2\Gamma ^n\epsilon _1 \bar{\theta }
                 \Gamma ^k\psi _n , \\ 
 Y_b^{\sp a} &=& -\bar{\theta }\Gamma ^n\epsilon _1 \bar{\epsilon }_2
                 \Gamma ^k\psi _n\hat{\omega }_{k\sp b}^{\sp a}
                 -\epsilon _1^{\sp \nu }\partial _{\nu }\Lambda _{2b}^{
                 \sp \sp a}
                 \nonumber \\
             & & +\frac{1}{144}
                 \bar{\theta }\Gamma ^k\epsilon _1\bar{\psi }_k(
                  \Gamma _{\sp b}^{a\sp rstu}\hat{F}_{rstu}+24\Gamma _{rs}
                  \hat{F}_{\sp b}^{a \sp rs})\epsilon _2 \nonumber \\
             & & -\frac{1}{18}\bar{\epsilon }_1\Gamma ^r\psi _w
                 \bar{\theta }\Gamma _{\sp b}^{a\sp wstu}\epsilon _2\hat{F}_{
                 rstu}-\frac{2}{3}\bar{\epsilon }_1\Gamma ^t\psi _k 
                 \bar{\theta }\Gamma ^{ku}\epsilon _2\hat{F}^a_{\sp btu}
                 \nonumber \\
             & & -\frac{1}{3}\bar{\theta }\Gamma ^{tu}\epsilon _2
                 \bar{\epsilon }_1\Gamma ^r\psi ^a\hat{F}_{rbtu}
                 -\frac{1}{3}\bar{\theta }\Gamma ^{tu}\epsilon _2
                 \bar{\epsilon }_1\Gamma ^s\psi _b\hat{F}^a_{\sp stu} 
                 \nonumber \\
             & & -2\bar{\theta }\Gamma ^{n}\epsilon _2e^{ka}e^l_{\sp b}
                 (-\bar{\epsilon }_1\Gamma _l\hat{D}_{[n}\psi _{k]}
                  +\bar{\epsilon }_1\Gamma _k\hat{D}_{[n}\psi _{l]}
                  +\bar{\epsilon }_1\Gamma _n\hat{D}_{[k}\psi _{l]})
                 \nonumber \\
             & & -\frac{1}{3}\bar{\theta }\Gamma ^{n}\epsilon _2
                  \bar{\epsilon }_1\Gamma ^{st}\psi _n\hat{F}^a_{
                  \sp bst} \nonumber \\
             & & -\frac{1}{108}(\bar{\theta }\Gamma _{\sp b}^{a\sp rstu}
                 \epsilon _2 +24\bar{\theta }\Gamma ^{tu}\epsilon _2
                 e^{ra}e^s_{\sp b})\bar{\epsilon }_1\Gamma _{[rs}^{
                 \sp \sp \sp wxy}\psi _t \hat{F}_{u]wxy}
                 \nonumber \\
             & & -\frac{1}{6}(\bar{\theta }\Gamma _{\sp b}^{a\sp rstu}
                 \epsilon _2 +24\bar{\theta }\Gamma ^{tu}\epsilon _2
                 e^{ra}e^s_{\sp b})\bar{\epsilon }_1\Gamma _{[rs}\hat{D}_{
                 t}\psi _{u]} \nonumber \\
             & & -\frac{1}{18}(\bar{\theta }\Gamma _{\sp b}^{a\sp rstu}
                 \epsilon _2 +24\bar{\theta }\Gamma ^{tu}\epsilon _2
                 e^{ra}e^s_{\sp b})\bar{\epsilon }_1\Gamma ^w\psi _{[r}
                 \hat{F}_{stu]w}.
\end{eqnarray}
 There are $\partial _m$ terms in $Y_b^{\sp a}$ . Thus $\Xi ^{k(2)}$ is 
 different from that in reference~\cite{wit98}.
 The other components are also different from that in 
reference~\cite{wit98}.

\section{Discussion}
Using 2nd-order formalism we obtained the same results 
as that was given in reference~\cite{wit98} up 
to obtained components, but at higher order it seems to be 
different from that in 1.5-order formalism. 
In order to obtain
superspace geometry which holds $\kappa $-invariance 2nd-order
formalism may be much simpler and more hopeful than 1.5-order formalism.
As seen in section 2, the condition for $\kappa $-invariance is given 
on torsion.
From the definition of torsion (\ref{tor}),
\begin{eqnarray}
 E_m^{\sp \beta }E_n^{\sp \gamma }
 T^a_{\sp \gamma \beta }= 2\partial _{[n}E_{m]}^{\sp \sp a}+2E_{[m}^{
                          \sp \sp b}\Omega _{n]b}^{\sp \sp \sp a},
\end{eqnarray}
at $\theta =0$ component,
\begin{eqnarray}
 \psi _m^{\sp \beta }\psi _n^{\sp \gamma }
 T^a_{\sp \gamma \beta }= 2\partial _{[n}e_{m]}^{\sp \sp a}-2e_{[m}^{
                          \sp \sp b}\hat{\omega}_{n]\sp b}^{\sp a}.
\end{eqnarray}
This is compatible with definition of spin connection (\ref{conn}) and
constraint for $\kappa $-invariance (\ref{kapcon}).

If we use 1.5-order formalism, the constraint (\ref{kapcon}) is not 
invariant under supersymmetry transformations.
Thus $\kappa $-invariance of action (\ref{mem}) will be realized in
very complicated form at higher order in anticommuting coordinates.

But, if we use 2nd-order formalism we can obtain superfields which hold 
the conditions (\ref{kapcon}). Thus for higher components in order to 
hold $\kappa $-invariance 2nd-order
formalism is expected to be more hopeful than 1.5-order formalism.

\section*{Acknowledgments}

I would like to thank Y.Matsuo for valuable suggestions, and thank
K. Hosomichi for valuable comments and discussions. 
\appendix
\section*{Appendix}
Our notations are almost same as that used in a text by 
J. Wess and J. Bagger~\cite{sg}.

\section{Indices}

We use Greek indices for spinorial components and Latin indices for
vector components. And we use former alphabet for the tangent space
indices and later for general coordinates indices: $a,b,c,...$ for tangent
vector indices and $k,l,m,...$ for general vector indices,
and  $\alpha ,\beta ,...$ for tangent spinorial indices and 
$\mu , \nu ,...$ for general spinorial indices.

Superspace coordinates $(x^m ,\theta ^{\mu })$ are designated 
$Z^M $ , where later capital Latin alphabet $M,N,..$ are collective 
designations for general coordinate indices. While former capital
Latin alphabet $A,B,..$ are collective designations for tangent
space indices.

\section{p-form superfield}

Vielbein is represented by $E^{\sp A}_{M}$ and its inverse is
$\tilde{E}^{\sp M}_{A}$ , which is defined as follows,

\begin{eqnarray}
 \tilde{E}^{\sp M}_{A}E^{\sp B}_{M}=\delta ^B _A , \nonumber \\
 E^{\sp A}_{N}\tilde{E}^{\sp M}_{A}=\delta ^M _N. 
\end{eqnarray}
We introduce p-form superfields as follows,

\begin{eqnarray}
  X               &\equiv & \frac{1}{p!} dz^{M_p}...dz^{M_1} 
                            X_{M_p ... M_1} \nonumber \\
                  &\equiv & \frac{1}{p!} E^{A_p}...E^{A_1} X_{A_p ... A_1}, \\
  X_{A_p ... A_1} &\equiv & \sum_{i=1}^{32} X_{A_p ... A_1}^{\sp \sp \sp (i)}.
\end{eqnarray}
 $X_{A_p ... A_1}^{\sp \sp \sp (i)}$ is component at i-th order in 
anticommuting coordinates.

\section{Convention}

We use the mostly plus metric; $\eta_{ab}\sim (-+...+)$ .
Symmetrization bracket $(\sp \sp )$ and antisymmetrization bracket
$[\sp \sp ]$ is defined as follows,

\begin{eqnarray}
   [M_1 ... M_N]   &=& \frac{1}{N!}( M_1 ... M_N \sp +
                               \mbox{antisymmetric terms} ), \nonumber \\
   (M_1 ... M_N)   &=& \frac{1}{N!}( M_1 ... M_N \sp +
                               \mbox{symmetric terms} ).
\end{eqnarray}
 
\section{Gamma matrices(11-dimensional)}

Since we use the Majorana representation, all components are real.

Gamma matrix $\Gamma^{a \sp \alpha}_{\sp \sp \sp \beta}$ is
defined as follows,

\begin{eqnarray}
 \{ \Gamma^{a} ,\Gamma^{b} \} = 2 \eta ^{ab}.
\end{eqnarray}

We lower the spinorial indices
by charge conjugation matrix $C_{\alpha \beta }$.
\begin{eqnarray}
  \bar{\psi}_{\beta} = \psi ^{\alpha }C_{\alpha \beta }, \nonumber \\
  \Gamma^a _{\sp \sp \alpha \beta}=C_{\alpha \gamma }
  \Gamma^{a \sp \gamma}_{\sp \sp \sp \beta} .
\end{eqnarray}
 The symmetric matrices are
\begin{eqnarray}
 \Gamma ^a ,\Gamma ^{a_1 a_2},\Gamma ^{a_1 ...a_5}, \Gamma ^{a_1 ...a_6},
 \Gamma ^{a_1 ...a_9},\Gamma ^{a_1 ...a_{10}},
\end{eqnarray}
 and antisymmetric matrices are 
\begin{eqnarray}
 C ,\Gamma ^{a_1 ...a_3},\Gamma ^{a_1 ...a_4}, \Gamma ^{a_1 ...a_7},
 \Gamma ^{a_1 ...a_8},\Gamma ^{a_1 ...a_{11}}.
\end{eqnarray}


\end{document}